\documentclass{PoS}
\usepackage{url}
\newcommand{\Process}{e^+e^-\rightarrow\chi\chi\gamma}
\definecolor{DESYorange}{RGB}{242,142,0}

\title{WIMP searches at the International Linear Collider}

\ShortTitle{WIMPs at the ILC}

\author{Keisuke Fujii\\
	High Energy Accelerator Research Organization, Japan\\
	E-mail: \email{keisuke.fujii@kek.jp}}
\author{\speaker{Moritz Habermehl}\\%\thanks{A footnote may follow.}
         Deutsches Elektron-Synchrotron, Germany\\
        E-mail: \email{moritz.habermehl@desy.de}}
\author{Jenny List\\
	Deutsches Elektronen-Synchrotron, Germany\\
	E-mail: \email{jenny.list@desy.de}}
\author{Shigeki Matsumoto\\
	Kavli Institute for the Physics and Mathematics of the Universe, Japan\\
	E-mail: \email{shigeki.matsumoto@ipmu.jp}}
\author{Tomohiko Tanabe\\
	International Center for Elementary Particle Physics, University of Tokyo, Japan\\
	E-mail: \email{tomohiko@icepp.s.u-tokyo.ac.jp}}

\abstract{Weakly Interacting Massive Particles (WIMPs, $\chi$) are candidates for Dark Matter. WIMP searches at lepton colliders are complementary to searches at hadron colliders and direct detection, since they directly probe the coupling to electrons which a priori is independent of the coupling to hadrons. Like at hadron colliders, WIMP pair production can be observed via an additional visible particle, in particular a photon from initial state radiation ($\Process$). With this technique WIMP masses to nearly half the centre-of-mass energy can be probed. Polarised beams are essential to reduce Standard Model backgrounds and to characterise the properties of the new particles in case a signal is discovered. Prospects for a mono-photon WIMP study at the International Linear Collider will be discussed in the context of EFT. In addition, detector requirements critical to this analysis are discussed.}

\FullConference{38th International Conference on High Energy Physics\\
		3-10 August 2016\\
		Chicago, USA}

\begin{document}

\section{Introduction}

The International Linear Collider (ILC) is a future electron-positron collider with a mature technology \cite{Behnke:2013xla}. 
%A conceptual sketch of the accelerator layout is shown in fig.\ref{fig:ILC}, left. 
Currently, a political decision in Japan is awaited. 
The centre-of-mass energy can be tuned between 250\,GeV and 500\,GeV and is upgradable to 1\,TeV. The instantaneous luminosity for $\sqrt{s}$ = 500\,GeV is 1.8$\times 10^{34}$\,cm$^{-2}$s$^{-1}$ which can be doubled to 3.6$\times 10^{34}$cm$^{-2}$s$^{-1}$ after a luminosity upgrade. Both the electron and position beams are foreseen to be polarised to at least $\pm$80\% and $\pm$30\%, respectively. The ILC will have one interaction region which will accommodate the two foreseen detectors in a push-pull scheme. The presented study is performed for the International Large Detector (ILD%, fig.\ref{fig:ILC}, right
) \cite{Behnke:2013lya}.

%\begin{figure}
% \begin{minipage}{0.65\textwidth}
%   \includegraphics[width=\textwidth]{../../Images/ILC.png}
% \end{minipage}
% \begin{minipage}{0.24\textwidth}
%   \includegraphics[width=\textwidth]{../../Images/ILD.png}
% \end{minipage}
%  \caption{Conceptual scetch of the ILC (left) and of ILD (right)}
%  \label{fig:ILC}
%\end{figure}

At colliders, WIMPs could be pair produced. Since the WIMPs do not leave signals in the detector an additional particle is required to detect the process, for example a photon from initial state radiation (ISR). We look for the signal process $\Process$ whose signature is a single photon in an otherwise "empty" detector. This approach is quasi model-independent. Due to the known initial state, the missing four-momentum can be calculated using two observables, namely the photon energy $E_\gamma$ and the photon polar angle $\theta_\gamma$. 

\begin{figure}
\begin{center}
  \begin{minipage}{0.23\textwidth}
  \includegraphics[width=\textwidth]{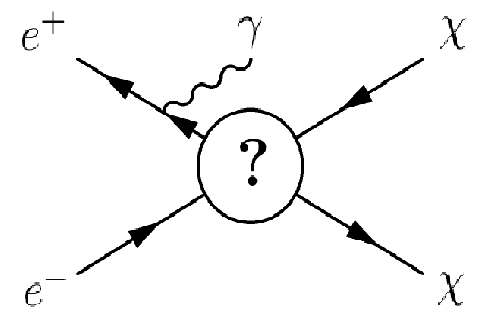}
 \end{minipage}
\hspace{1cm}
 \begin{minipage}{0.23\textwidth}
   \includegraphics[width=\textwidth]{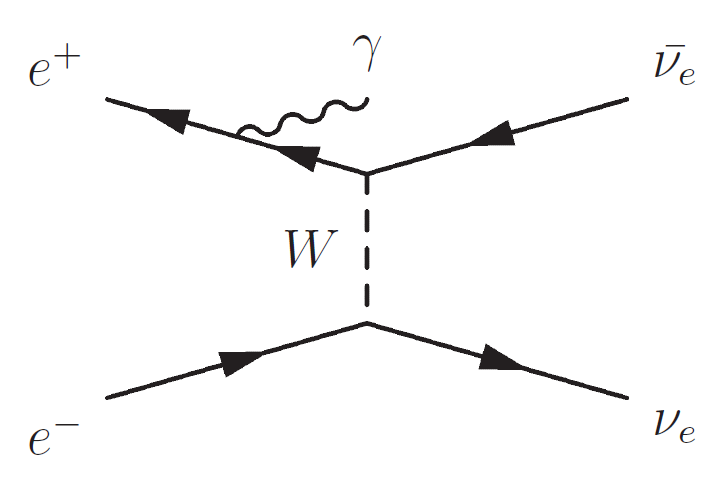}
 \end{minipage}
\hspace{1cm}
 \begin{minipage}{0.195\textwidth}
   \includegraphics[width=\textwidth]{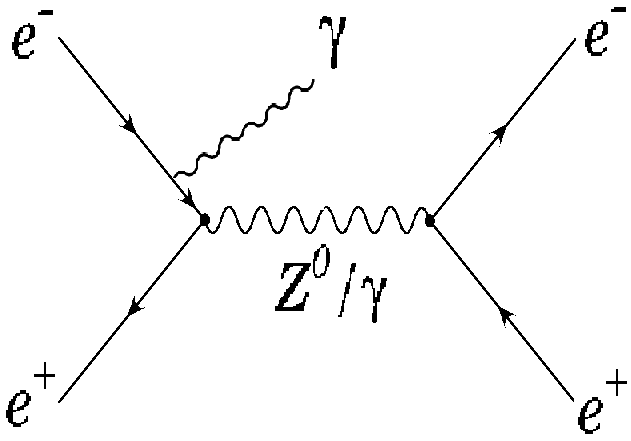}
 \end{minipage}
\end{center}
  \caption{Pseudo-Feynman diagram for the signal process and example Feynman diagrams for the two main background processes: radiative neutrino pair production and Bhabha scattering.}
  \label{fig:feynman}
\end{figure}

The two main background processes are neutrino pair production and Bhabha scattering, both with an associated photon from initial state radiation, or in the latter case also from final state radiation (see fig.\ref{fig:feynman}). The neutrino background is irreducible, but can be enhanced or suppressed by changing the polarisation combination. Bhabha scattering has a huge cross section and mimics the signal if both leptons escape undetected. For the suppression of this background process the best possible hermeticity in the forward region of the detector is required. 

\section{Motivation for an ILC Simulation Study}

The theoretical framework used in this analysis are effective operators, where the underlying idea is to classify the WIMP based on its quantum numbers (spin and weak isospin) and the mediator by its spin and construct the minimal effective Lagrangian. The only parameter that remains, $\Lambda$, can be called the energy scale of new physics and is a function of the mediator mass and the coupling to the fermions $g_f$ and the coupling to the WIMPs $g_\chi$: $\Lambda = M_{mediator} / \sqrt{g_fg_\chi}$. 

\begin{figure}
\begin{minipage}{0.35\textwidth}
 \includegraphics[width=\textwidth]{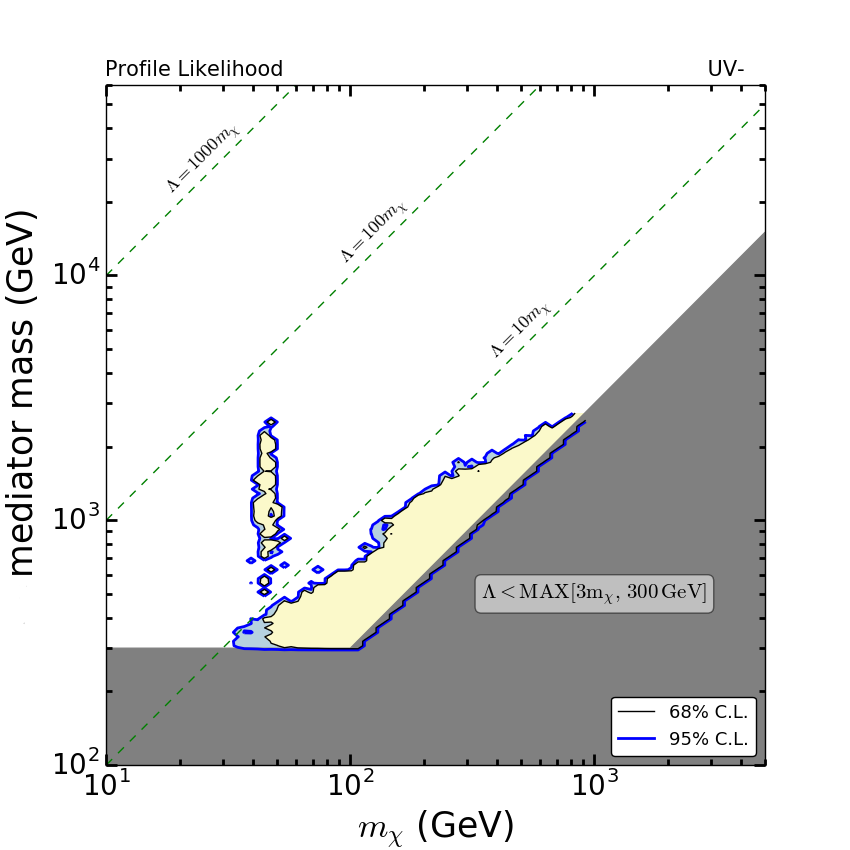}
\end{minipage}
\hspace{0.05\textwidth}
\begin{minipage}{0.6\textwidth}
   \caption[]{The parameter space which will not be covered by dark matter experiments by the time the ILC will be running is shown in yellow (68\% confidence level) and blue (95\% C.L.). The computations are based on extrapolations of current and future experiments assuming that no WIMP signal is detected. The grey area reflects the parameter space in which the approach of effective operators is not valid: $\Lambda$ has to be larger than three times the WIMP mass and above 300\,GeV. \cite{Matsumoto:2016hbs}}
\label{fig:EFT_shig}
\end{minipage}
\end{figure}

In such a framework, considering the full Lagrangian, a likelihood analysis of data of existing experiments, together with extrapolations of expected exclusion limits at the time the ILC is running is being performed. Figure \ref{fig:EFT_shig} shows the surviving region assuming that no WIMP signal is detected, for the example of a singlet-like fermion WIMP \cite{Matsumoto:2016hbs}. Data from the following experiments are considered: from the Planck satellite, from the direct detection experiments PICO-2L, LUX and XENON100, from collider searches at LEP and LHC and from future experiments like LZ and PICO250. Here, the couplings are tested in the range [-1,1]. Above the grey area this simplified model reproduces the results of effective operators. The yellow region shows the parameter space which will not be explored by other experiments before the ILC starts. 

\section{Modelling of Signal and Background}

The signal definition in this analysis comprises three requirements on the photon: a minimum energy of 10\,GeV, a maximum energy of 220\,GeV and $|\cos\theta_\gamma|<$ 0.98. The upper cut on the energy is applied to avoid the large background rates in the region close to the radiative return to the Z boson, which corresponds to E$_\gamma$ = 242\,GeV for $\sqrt{s}$ = 500\,GeV. The angle is restricted to the parts of the detector in which the tracking performance guarantees that photons can be distinguished from electrons and positrons.
%Additionaly, with the hard photon defined by the combination of the cuts, the system of the colliding electron and positron has a sizeable transverse momentum. Hence the peak of the distribution of the transverse momentum of the leptons of Bhabha scattering events is also shifted to higher values and allow their detection and hence these events can be distinguished them from signal-like events.

The events are generated using WHIZARD version 2.2.8 \cite{Kilian:2007gr}, with the matrix element generator O'Mega \cite{Moretti:2001zz}. Polarised beams are included as well as the beam energy spectrum. The generated background processes are neutrino pairs plus several photons and for the Bhabha scattering electron-positron pairs plus several photons. The signal events $\chi\chi\gamma$ are obtained by reweighting the $\nu\bar{\nu}\gamma$ events using the differential cross section  formulas expressed in terms of the WIMP mass and spin.
%according to the WIMP parameters, like mass, spin etc.

For modelling the photons, WHIZARD offers an ISR parametrisation that comprises all orders of soft-collinear photons and the first three orders of hard-collinear photons. With this the cross sections of the considered processes are calculated with high accuracy. However, a realistic distribution of the photon polar angle is obtained by including the photons in the matrix element. By doing so, double counting of photons is avoided. Both approaches are combined by generating the events with the photons in the matrix element and reweighting the cross section to the one with the ISR parametrisation.

The events are simulated in a Geant4 based simulation of the full ILD detector model presented in the Technical Design Report \cite{Behnke:2013lya}. 

\section{Results}
\subsection{Higher Sensitivity with Improved Bhabha Rejection}
\label{sec:newL}

\begin{table}[h]
\begin{minipage}{0.52\textwidth}
   \begin{tabular}{ | l | r | r | }
    \hline
    $e^+e^-\gamma$ & \hphantom{new analys}\cite{Bartels:2012ex} & new analysis \\
    \hline
    $p_T$ &  21.1\hphantom{9.}\%  &  26.1\hphantom{2.}\% \\
    $E_{vis}$  &  16.0\hphantom{9.}\% & 1.9\hphantom{2.}\% \\
    BeamCal\hphantom{..} & 0.29\hphantom{.}\%  & 0.02\hphantom{.}\% \\
    \hline
  \end{tabular} 
\end{minipage}
\begin{minipage}{0.48\textwidth}
  \caption{Fractions of signal-like Bhabha scattering events surviving the three conditions for an "empty" detector. The central column shows the latest full analysis \cite{Bartels:2012ex} and the right one our update.}
\label{tab:cutflow}
\end{minipage}
\end{table}

Signal-like events with only very little detector activity besides the photon are selected by requiring the following three criteria. There must be no tracks with a transverse momentum exceeding 3\,GeV. The visible energy, excluding that of the photon, has to be smaller than 20\,GeV. Finally, there must be no electrons and positrons detected in the forward region, or more technically speaking: no clusters may be reconstruced in the forward detector BeamCal.

Table \ref{tab:cutflow} shows the percentage of Bhabha scattering events surviving the three conditions, for the full analysis done in 2013 \cite{Bartels:2012ex} (central column) together with our update of the analysis. In the new analysis the Bhabha background is suppressed to a sub-permille level with respect to the signal definition, which is approximatly fifteen times more effective compared to the last analysis.
The requirement of a low visible energy leads to an eight times higher rejection, which is due to improvements in particle flow and photon reconstruction.
%The improvements in particle flow and photon reconstruction since the last analysis leads to a eight times better rejection with the requirement of a low visible energy. 
With an optimised BeamCal reconstruction algorithm \cite{Sailer:2227265}, an additional factor of two is achieved. This shows that both detector resolution and detailed reconstruction are crucial for a reliable performance estimate. 

\begin{figure}
 %\begin{minipage}{0.3\textwidth}
 % \includegraphics[width=\textwidth]{../../Images/old_vs_new_2_valid.png}
 %\end{minipage}
 \begin{minipage}{0.4\textwidth}
  \includegraphics[width=\textwidth]{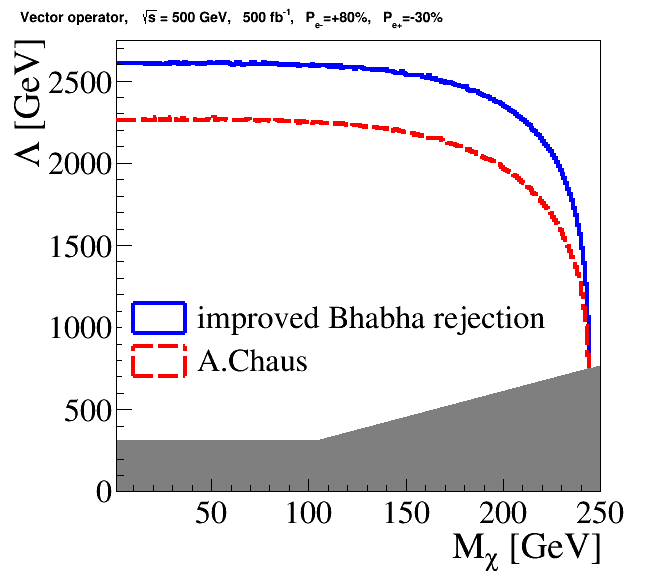}
 \end{minipage}
\begin{minipage}{0.6\textwidth}
   \caption{The area below the curves shows the excluded $\Lambda$ values as a function of the WIMP mass for a vector-like fermion WIMP and a vector-like operator, $\sqrt{s}$ = 500\,GeV, 500\,fb$^{-1}$, right-handed electrons and left-handed positrons. The red dashed curve is for the Bhabha scattering background suppression shown in the central column of the table and the blue solid line is for the improved background rejection as shown in the right column of the table.}
  \label{fig:newresults}
\end{minipage}

\end{figure}

Figure \ref{fig:newresults} shows a comparison between the previous sensitivity \cite{chaus2014} which is based on the background rejection efficiency of \cite{Bartels:2012ex} (red dashed line) and our update (blue solid line). For right-handed electrons and left-handed positrons the level of the exclusion limit can be improved by up to 300\,GeV in $\Lambda$, for a centre-of-mass energy of 500\,GeV and an integrated luminosity of 500\,fb$^{-1}$. 
%The sensitivity is improved by 6\% even when no beam polarisation is assumed.

\subsection{Impact of the Forward Detector Design}

High numbers of electron-positron pairs created from beamstrahlung hit the inner part of BeamCal. As a consequence, the reconstruction efficiency of particles coming from the hard interaction decreases dramatically at very low polar angles $\theta$. The identification of Bhabha scattering events hence strongly depends on the design of the forward region. 

Because the ILC detectors share the same interaction region, both are required to have the same focal distance of the final quadrupole magnet (L*). In order to equalise that length for both detectors, the forward region of the ILD detector is currently being redesigned \cite{cr2}. BeamCal has to be moved closer to the interaction point by 40\,cm.
While the detailed redesign and the implementation in the ILD simulation is ongoing, we estimate the impact in the following way:
Firstly, we assume that area in BCal polluted by the pairs is the same for all considered distances. Secondly, the lepton reconstruction efficiency is approximated to be 100\% for polar angles above an effective angle $\theta_{eff}$ and 0\% below. The number of events with both leptons not being reconstructed using the full detector simulation corresponds to an effective angle of approximately 15.5 mrad.
With these assumptions, the hermeticity in the forward region depends purely geometrically on the distance of BeamCal to the interaction point.
This means that the effective polar angle grows linearly with the distance by which BeamCal is moved in towards the interaction point. 

\begin{figure}
\begin{minipage}{0.35\textwidth}
 \includegraphics[width=\textwidth]{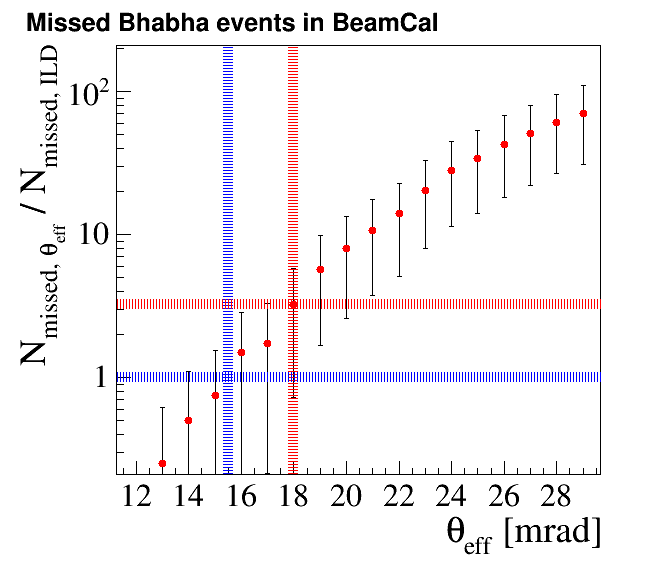}
\end{minipage}
\begin{minipage}{0.65\textwidth}
  \caption{The number of Bhabha scattering events with both leptons lying within an effective polar angle, normalised to the number of events with both leptons not being reconstructed using the full detector simulation, which corresponds to $\theta_{eff}\approx$\,15.5 mrad. The blue bands show the situation for the current detector design and the red bands indicate the BeamCal which is moved 40\,cm closer to the interaction point. We estimate that the Bhabha background will increase by a factor of three to four.}
  \label{fig:BC_bars}
\end{minipage}
 
\end{figure}

Figure \ref{fig:BC_bars} shows how the number of Bhabha scattering events with both leptons not being reconstructed changes as a function of the effective polar angle. With the new position of the subdetector, the Bhabha backgrond would be three to four times higher. With this higher background level the improvement presented in section \ref{sec:newL} would be partially lost. 
%A full ILD update is underway. 

\subsection{Sensitivity in Different Operation Scenarios}

\begin{figure}
 \begin{minipage}{0.63\textwidth}
  \includegraphics[width=.49\textwidth]{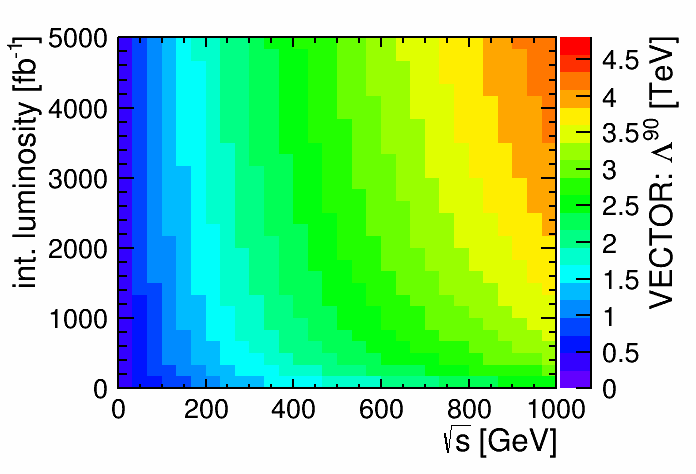}
\includegraphics[width=.49\textwidth]{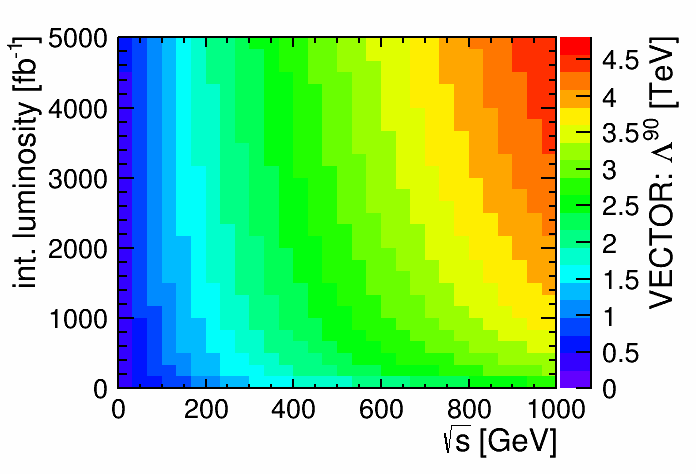}
  \caption{Extrapolation of the exclusion limits from the full simulation to the full range of ILC centre-of-mass energies and different integrated luminosities, for fractions of 22.5\% (left) and 40\% of the data collected with right-handed electrons and left-handed positrons.}
  \label{fig:extrapol}
\end{minipage}
 %\begin{minipage}{0.3\textwidth}
  %\includegraphics[width=\textwidth]{../../Images/V_67_scale48.png}
  %\caption{Extrapolation of the exclusion limits from the full simulation to the full range of ILC centre-of-mass energies and different integrated luminosities.}
  %\label{fig:extrapol}
%\end{minipage}
\hspace{0.05\textwidth}
 \begin{minipage}{0.27\textwidth}
  \includegraphics[width=\textwidth]{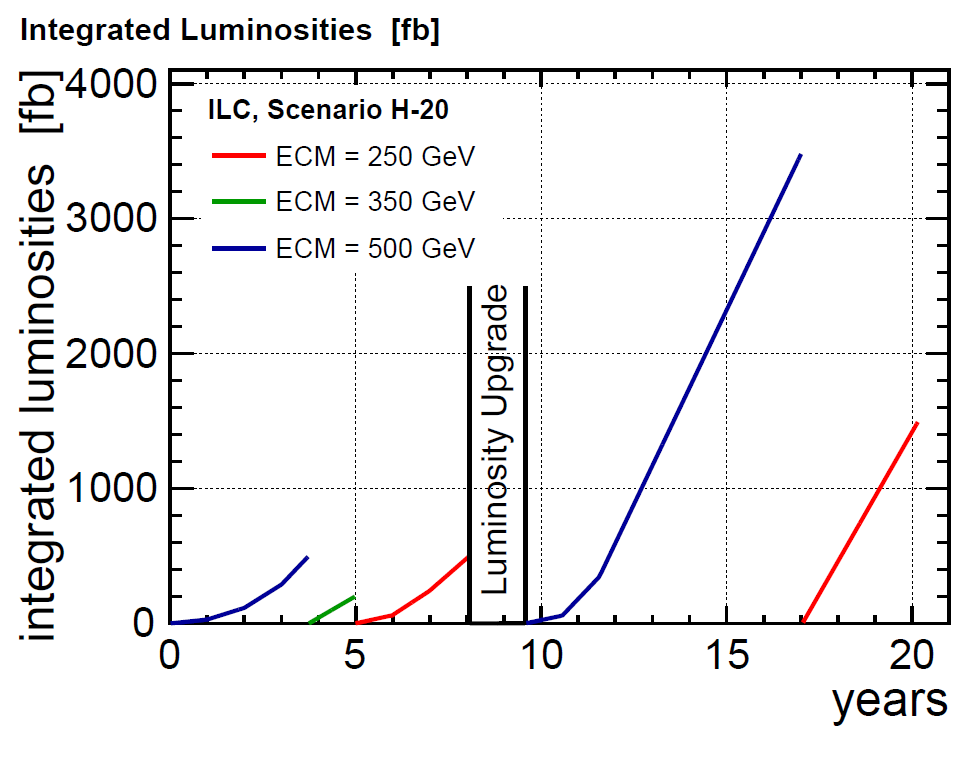}
\caption{The H-20 running scenario which is one of the possible 20 years programmes for the ILC\cite{Barklow:2015tja}.}
\label{fig:H20}
 \end{minipage}

\end{figure}

Based on the results of the previous study \cite{Bartels:2012ex} performed at $\sqrt{s}$ = 500\,GeV and 500\,fb$^{-1}$ we developed a scheme to extrapolate the results to other energies and integrated luminosities. This extrapolation is restricted to WIMP masses below 100\,GeV, where the sensitivity does not depend on the WIMP mass. This study allows to give estimates of the sensitivity for different time scales and different running scenarios, i.e. how much integrated luminosity is collected at which centre-of-mass energy, in which order this occurs and how the integrated luminosity is distributed between the different polarisation combinations (fig. \ref{fig:extrapol}). 

The running scenario H-20 ~\cite{Barklow:2015tja} starts with $\sqrt{s}$ = 500\,GeV at which initially 500\,fb$^{-1}$ are collected (fig.\ref{fig:H20}). After the first four years an exclusion limit of $\Lambda\approx$ 2.5\,TeV is reached for a vector-like fermion WIMP and a vector-like operator. When ignoring the contributions of the other centre-of-mass energies, this exclusion limit grows during a second run at $\sqrt{s}$ = 500\,GeV after a luminosity upgrade. After the full ILC programme the runs at $\sqrt{s}$ = 500\,GeV will have contributed to an exclusion limit of $\Lambda\approx$ 3\,TeV. With an upgrade to $\sqrt{s}$ = 1\,TeV the reachable $\Lambda$ is approximately 4.5\,TeV. 

The sensitivity strongly depends on the fraction of the integrated luminosity collected with right-handed electrons and left-handed positrons for which the neutrino background is strongly suppressed. The rather large fraction in H20 (40\%) is clearly favoured over 22.5\% (compare fig.\ref{fig:extrapol}).

A full update of the whole analysis to the new detector performance for all types of WIMPs is underway.

\bibliographystyle{ieeetr}
\bibliography{Proceeding_WIMPs_V6}

%\begin{thebibliography}{99}

%\bibitem[1]{Shig} Matsumoto et al.: "Effective Theory of WIMP Dark Matter supplemented by Simplified Models: Singlet-like Majorana fermion case", \textit{arXiv:1604.02230}
%\bibitem[2]{TDR} Behnke et al., "The International Linear Collider Technical Design Report - Volume 1: Executive Summary", \textit{arXiv:1306.6328}
%\bibitem[3]{ILD} Behnke et al., "The International Linear Collider Technical Design Report - Volume 4: Detectors", \textit{arXiv:1306.6329}
%\bibitem[4]{whizard} W. Kilian, T. Ohl, J. Reuter, "WHIZARD: Simulating Multi-Particle Processes at LHC and ILC", \textit{arXiv:0708.4233}
%\bibitem[5]{whizard2} M. Moretti, T. Ohl, J. Reuter, "O'Mega: An Optimizing Matrix Element Generator", \textit{arXiv:hep-ph/0102195}
%\bibitem[6]{CB} Christoph Bartels, Mikael Berggren and Jenny List, "Characterising WIMPs at a future e$^+$e$^-$ Linear Collider", \textit{arXiv:1206.6639}
%\bibitem[7]{CBDA} Christoph Bartels, Phd thesis, 2011, DESY-THESIS-2011-034
%\bibitem[8]{AC} Andrii Chaus, PhD thesis, Universit\'e Paris-sud 11, 2014PA112300
%\bibitem[9]{run} Barklow et al., "ILC Operating Scenarios", \textit{arXiv:1506.07830}
%\bibitem[10]{clic} CLICdp-Note-2016-005, 25 October 2016
%\bibitem[11]{cr2} Benno List Proceedings of IPAC2016, Busan, Korea, THPOR026

%\end{thebibliography}

\end{document}